\documentclass{article}
\usepackage{spconf,amsmath,graphicx}
\usepackage{siunitx}        
\usepackage{booktabs}       
\usepackage{caption}
\usepackage[acronym,nowarn,hyperfirst=false]{glossaries}

\setkeys{glslink}{hyper=false}
\usepackage[backend=bibtex,style=ieee,natbib=true]{biblatex} 
\title{Low-Dimensional Bottleneck Features for On-Device Continuous Speech Recognition}
\name{David B. Ramsay$^{\star}$\sthanks{The first author performed this work while at Google AI.} \qquad Kevin Kilgour$^{\dagger}$ \qquad Dominik Roblek$^{\dagger}$ \qquad Matthew Sharifi$^{\dagger}$}

	\address{$^{\star}$ MIT Media Laboratory \\
	    $^{\dagger}$ Google AI}

\renewcommand*{\CustomAcronymFields}{%
  name={\the\glsshorttok},%
  description={\the\glslongtok},%
}

\SetCustomStyle

\newacronym{DSP}{DSP}{digital signal processor}
\newacronym{BNF}{BNF}{bottleneck feature}
\newacronym{LVCSR}{LVCSR}{large-vocabulary continuous speech recognition}
\newacronym{WER}{WER}{word error rate}
\newacronym{LAS}{LAS}{Listen, Attend, Spell}
\newacronym{NN}{NN}{neural networks}
\newacronym{ASR}{ASR}{automatic speech recognition}
\newacronym{HMM}{HMM}{hidden markov model}
\newacronym{BW}{BW}{bandwidth}
\newacronym{BN}{BN}{bottleneck}
\newacronym{ReLU}{ReLU}{restricted linear unit}
\newacronym{QMF}{QM-features}{quantized mel features}

\begin{document}
%
\maketitle

\begin{abstract}
Low power \glspl{DSP} typically have a very limited amount of memory in which to cache data. In this paper we develop efficient \gls{BNF} extractors that can be run on a \gls{DSP}, and retrain a baseline \gls{LVCSR} system to use these \glspl{BNF} with only a minimal loss of accuracy. The small \glspl{BNF} allow the \gls{DSP} chip to cache more audio features while the main application processor is suspended, thereby reducing the overall battery usage.  Our presented system is able to reduce the footprint of standard, fixed point \gls{DSP} spectral features by a factor of 10 without any loss in \gls{WER} and by a factor of 64 with only a \SI{5.8}{\%} relative increase in \gls{WER}.
\end{abstract}
\begin{keywords}
bottleneck features, large vocabulary continuous speech recognition, low-power deep learning, mobile
\end{keywords}

\section{Introduction}
\glsreset{LVCSR}
\glsreset{BNF}
\glsreset{DSP}

\Gls{LVCSR} can be used to extract rich context about a user's interests, intents, and state. If run on a mobile device, this has the potential to revolutionize the quality of on-device services they interact with. In order for this to become practical, hardware-level optimization is required to preserve the battery life of portable devices. 

In this paper, we present a new \gls{LVCSR} model architecture that takes advantage of a low-power, fixed point, always-on \gls{DSP} to significantly reduce power consumption. Our goal is to use the \gls{DSP} to optimally compress incoming speech into its \glspl{BNF} representation which is cached for as long a period as possible. By increasing the amount of cached input, we reduce the wake-up frequency of the device's main processor, which is used to complete the inference.

We start with a state-of-the-art \gls{LAS} end-to-end \gls{ASR} model, and effectively split its encoder across the \gls{DSP} and the main processor.  Hardware optimization across the \gls{DSP} and main processor has been successfully leveraged in the past to cache features for similar low-power services \cite{gfeller2017now}, though this is the first time that a \gls{DSP} has been used to compute the initial layers in the primary inference model.  
This leads to a significant increase in the amount of audio we can cache, with minimal impact to the model's overall \gls{WER}. Furthermore, as a purely on-device model, this design preserves user privacy as well as battery life. The topology is an important step towards practical \gls{LVCSR} in highly power-constrained contexts.

\begin{figure*}[th]
\centering
\includegraphics[width=.8\linewidth]{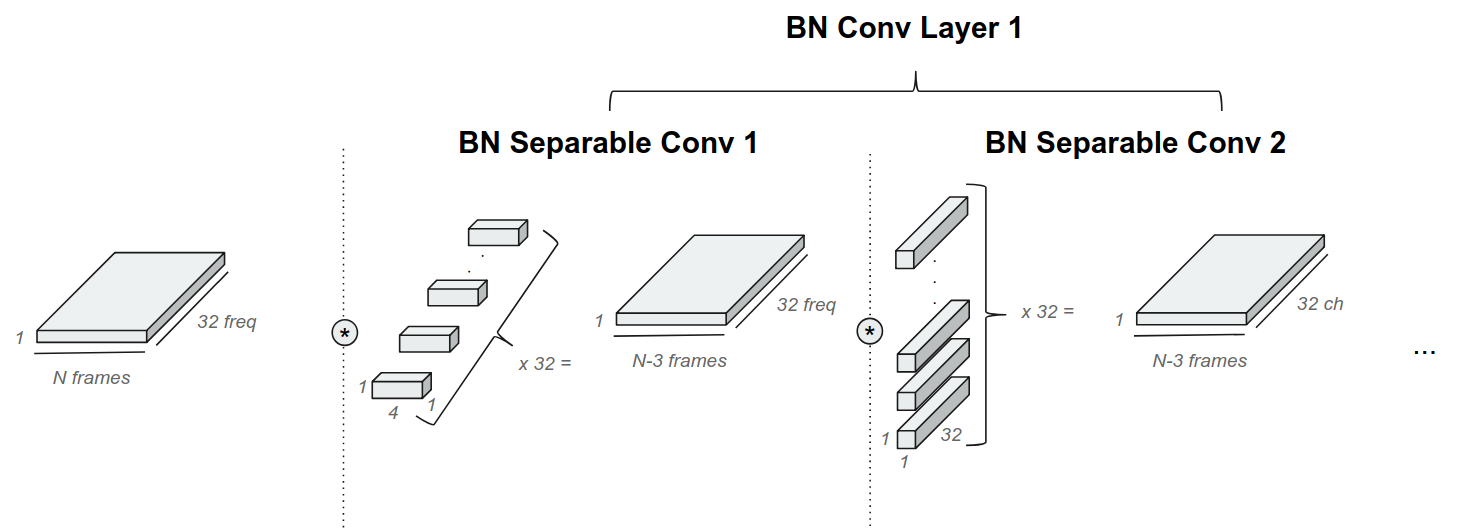}
\captionsetup{width=\linewidth}
\caption{\textit{The default configuration of a bottleneck layer running on the \gls{DSP}; here we see a kernel size of 4 applied in a frequency separable way, followed by one frequency kernel per output channel. These two convolutions are considered as a single 'layer'.}}
\label{bn_structure}
\vspace{-0.34cm}
\end{figure*}

\section{Related Work}
Fully end-to-end \gls{LVCSR} are emerging as the state-of-the-art \cite{chiu2018state}, equalling and even surpassing the performance of standard connectionist temporal classification \cite{graves2012connectionist} models. The core architecture for these end-to-end models, called Listen, Attend, and Spell \cite{chan2016listen}, contains three major subgraphs - an encoder, an attention mechanism, and a decoder.  Since their proposal in 2015, there has been a substantial amount of work done to optimize these models for on-device use \cite{prabhavalkar2016compression, pang2018compression}, including weight matrix factorization, pruning, and model distillation.  Due to these improvements, it is now possible to run a state-of-the-art \gls{LVCSR} model on a mobile device's core processor (at a high power cost).

For the traditional \gls{HMM}-based systems that predate \gls{LAS} architectures, \gls{NN} had been heavily used as part of a traditional \gls{ASR} acoustic model. \citet{vesely2011convolutive} show that convolutional bottleneck compression improves system performance in such setups. Typically, these compressed representations are concatenated with small time-window features to provide 'context'.

Additionally, small \gls{HMM}-based keyword spotters have been successfully optimized across a \gls{DSP} and main processor. \citet{shah2018fixed} propose a model which introduces \SI{5}- and \SI{6}{bit} weight quantization for a reduced memory footprint without a significant reduction in accuracy.  Although these models have different architectures and applications, their use of convolutional bottleneck features and fixed-point network quantization inform our architecture.

\citet{shah2018fixed}, \citet{gfeller2017now} introduce a split across a fixed-point \gls{DSP} and a main processor motivated by power optimization. A quantized, two-stage, separable convolutional layer running on the \gls{DSP} forms the basis of their music detector. We use the same layer structure in our \gls{DSP} implementation.

The previously mentioned approaches do not attempt to compress audio features before caching, but there are other analyses of the trade-off between feature caching and power savings in the literature. In \citet{priyantha2011littlerock} and \citet{priyantha2010eers}, empirical power consumption drops from \SI{700}{mW} to \SI{25}{mW} as data is cached \SI{50}{x} longer for a pedometer application.  Measurements of \citet{gfeller2017now} indicate a full \SI{25}{\%}-\SI{50}{\%} of the power cost at inference time is due to fixed wakeup and sleep overhead.  Our goal is to significantly reduce this fixed power cost.

\section{Feature Substitution}

State-of-the-art results are reported in \citet{chiu2018state} with a very large, proprietary corpus. In this paper, we use the Librispeech~100 corpus to train our model \cite{panayotov2015librispeech}. \citet{chiu2018state} report a \gls{WER} of \SI{4.1}{\%} with over 12,500 hours of training data; the same model trained on 100 hours of Librispeech data gives a \gls{WER} of \SI{21.8}{\%}, which we use as the baseline for all further evaluation.

The model from \citet{chiu2018state} is capable of running on a phone using 80-dimensional, \SI{32}{bit} floating point mel spectrum audio features sampled in \SI{25}{ms} windows every \SI{10}{ms}. These features capture a maximum frequency of \SI{7.8}{kHz} and are stacked with delta and double delta features, resulting in an 80 x 3 input vector at each timestep.  We replace these features with \textit{\gls{QMF}} that are compact, simple to calculate, and currently in use by other services running on the \gls{DSP}.

\gls{QMF} are log-mel based with a \SI{16}{bit} fixed point representation. We use a default, narrow-band frequency representation that only captures up to \SI{3.8}{Hz} over 32 bins. We test the effect of reducing the bandwidth by simply using fewer log-mel bins. Sampling rate and window size are constant across test input features and, for each case, we train an end-to-end model. The results of training a state-of-the-art \gls{LAS} model with different input representations which can be calculated and cached on the \gls{DSP} can be seen in Table~\ref{feature_table}.

\begin{table*}[t]
  \label{features}
  \centering
  \begin{tabular}{lllll}
    \toprule
		Model Input & Input Dims & Feature Type & \gls{WER} (\%) & \gls{BW} (kbps) \\ 
 	\midrule
    
\textit{\SI{16}{kHz} \SI{16}{bit} raw PCM audio} & -- & -- & -- & \textit{256} \\
\textbf{Baseline LAS Model} & 80 x 3 & Mel, +$\Delta$ +$\Delta\Delta$ & \textbf{21.79} & \textbf{768} \\
Standard \gls{QMF} + Deltas & 32 x 3 & Mel, +$\Delta$ +$\Delta\Delta$ & 22.42 & 154 \\
\textbf{Standard \gls{QMF}} & 32 x 1 & Mel & \textbf{22.62} & \textbf{51.2} \\
3/4 \gls{BW} \gls{QMF} & 24 x 1 & Mel & 22.80 & 38.4 \\
1/2 \gls{BW} \gls{QMF} & 16 x 1 & Mel & 22.97 & 25.6 \\
1/4 \gls{BW} \gls{QMF} & 8 x 1 & Mel & 24.52 & 12.8 \\
    \bottomrule
  \end{tabular}
    \caption{Comparison of model performance with smaller feature representations.}
  \label{feature_table}
  \vspace{-0.34cm}
\end{table*}

The results indicate that the baseline model, whose features have not previously been optimized, has a heavily redundant input representation, requiring three times the \gls{BW} of the raw audio after delta stacking.  We are able to significantly reduce the input \gls{BW} (and, by extension, the amount of computation in the initial \gls{LAS} layers) without severely affecting the model's \gls{WER}.  

Delta- and double delta- feature stacking do not have a large effect relative to their \SI{3}{x} increase in size; thus we will take the \textit{standard} \SI{32}{bin} \gls{QMF} input as our starting point for further exploration.  Though we see an incremental trade-off between \gls{BW} and \gls{WER} for smaller raw feature representations, we will use the full \SI{32}{bin}  \gls{QMF} as an input to our compressived bottleneck layers in an attempt to preserve \gls{WER} while reducing the \gls{BW} even more drastically.  

\begin{figure*}[t]
\centering
\includegraphics[width=0.49\linewidth]{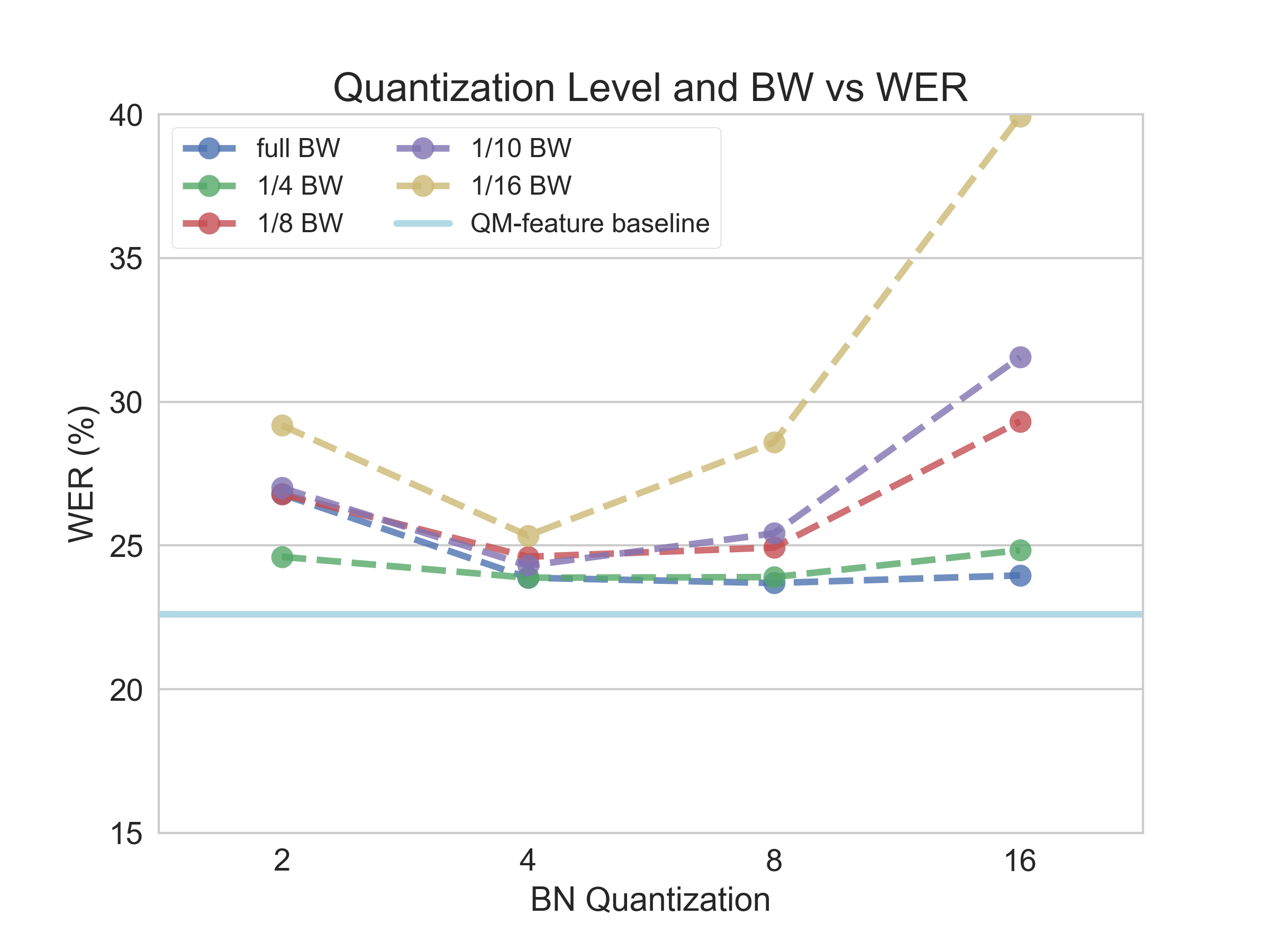}
\includegraphics[width=0.49\linewidth]{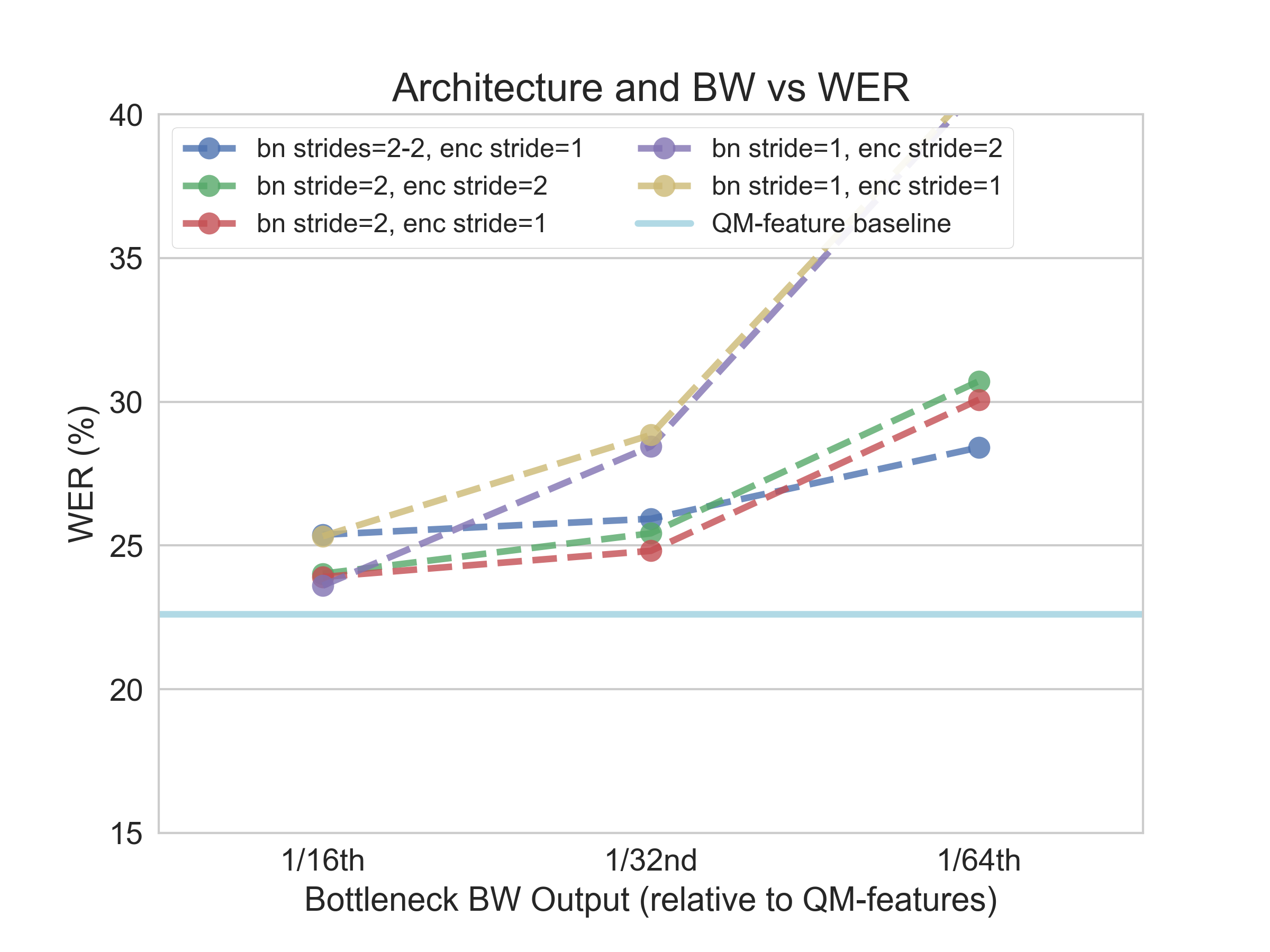}
\captionsetup{width=.9\linewidth}
\caption{The left plot uses a bottleneck feature extractor with a single hidden layer in which the output layer dimension and quantization level were modified to give a certain bandwidth output (relative to the standard \SI{32} dimensional \SI{16}{bit} \gls{QMF}). We see a trend towards 4-bit quantization, especially at high compression levels.  The right plot shows the performance of various architectures (different bottleneck and encoder depths/strides and \gls{BNF} dimension) at 4-bit quantization, plotted against bandwidth. As more drastic compression is demanded, shifting the stride to before the \glspl{BNF} improves performance, which is similar to reducing the frame rate in more traditional models \cite{Pundak45555}.}
\label{bn_results}
\vspace{-0.34cm}
\end{figure*}

\section{Bottleneck Feature Extraction}


Our model uses the convolutional structure outlined by \citet{gfeller2017now}. The structure of a single layer is shown in Figure~\ref{bn_structure}.  These simple, separable convolutional layers have been optimized for the \gls{DSP}. Besides minimal computation, all layer weights and intermediate representations are quantized to \SI{8}{bits}. \SI{32}{bit} biases, batch normalization \cite{ioffe2015batch}, and a \gls{ReLU} activation function are included after the second, 1-D separable convolution.

To explore the space of bottleneck architectures, we parameterized this architecture along the following axes: output dimension size, output quantization level, convolutional stride (in time), kernel size, and the number of layers in the bottleneck network. The first three axes have the potential to reduce the \gls{BW} of the resulting bottleneck, while the latter two axes are relevant to the size of the resulting model. Reducing the output dimension size is equivalent to reducing the size of the bottleneck layer and can result in a proportional reduction in \gls{BW}. The output quantization level affects how many bits are saved for each of the values in the output, and will also result in a proportional reduction in \gls{BW}. Increasing the stride could exponentially decrease the \gls{BW}, for example, by doubling the stride we generate outputs only half as often. 

These changes in input lead to a necessary modification of the initial two convolutional layers of the \gls{LAS} encoder, which are designed with 3x3 time-frequency kernels and strides of 2.  We replace these (by default) with a 3x1 time kernel along the flattened and modified frequency axis.  We also vary the number of initial encoder layers and strides in our analysis.

\section{Results}

\begin{table*}[t]
  \centering
  \resizebox{0.95\textwidth}{!}{%
  \begin{tabular}{lccccc}
    \toprule
		Model & $\#$ \gls{BNF} Extractor Weights & $\Delta$ \gls{LAS} Encoder Weights &  Total Stride\footnotemark & \gls{BW} (kbps) & WER (\%) \\ 
 	\midrule
\textit{16kHz 16-bit Raw PCM Audio} & -- & -- &  -- & \textit{256} & --\\
Baseline LAS Model & -- & \textit{0 (0)} & 4 & 768 & 21.79 \\
Standard \gls{QMF} & \textit{0 (0)} & -3,072 (-98KB) & 4 & 51.2 & 22.62 \\ [0.2cm]
Best \char`\~1/10 \gls{BW}. \gls{BNF} Model, $\nabla$ & 512 (4KB) & -8,064 (-258KB) & 1 & 4.8 & 22.44 \\
Best \char`\~1/20 \gls{BW}. \gls{BNF} Model, $\nabla$ & 512 (4KB) & -8,064 (-258KB) & 2 & 2.4 & 23.55 \\
Best 1/32 \gls{BW}. \gls{BNF} Model, $\nabla$ & 384 (3KB) & -8,448 (-270KB) & 2 & 1.6 & 24.81  \\ [0.2cm]
Best 1/16 \gls{BW}. \gls{BNF} Model & 640 (5KB) & -7,680 (-246KB) & 4 & 3.2 & 24.02  \\
1/32 \gls{BW}. \gls{BNF} Model & 384 (3KB) & -8,448 (-270KB) & 4 & 1.6 & 25.42  \\
Best 1/64 \gls{BW}. \gls{BNF} Model & 1536 (123KB) & -8,448 (-270KB) & 4 & 0.8 & 28.41 \\
\bottomrule
  \end{tabular}}
\caption{ Selection of best performing models for different bandwidths.}
\label{results_table}
\vspace{-0.34cm}
\end{table*}

\begin{figure}[t]
\centering
\includegraphics[width=0.98\linewidth]{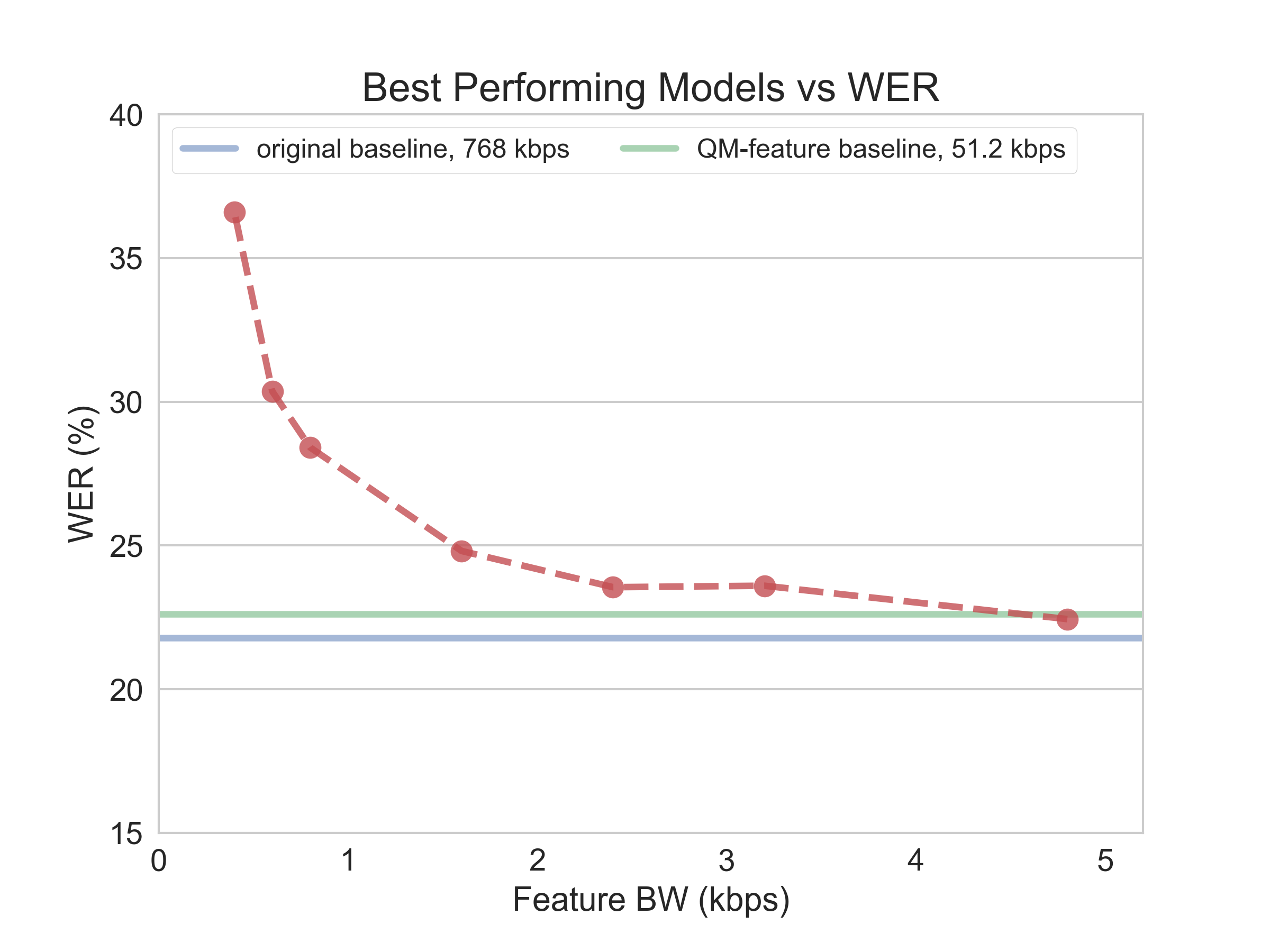}

\captionsetup{width=.9\linewidth}
\caption{Best performing model vs bandwidth.  We see a good trade-off around \SI{2}{kbps}.}
\label{final_results}
\vspace{-0.34cm}
\end{figure}





Initial results are based on freezing the \gls{BN} extractor and encoder layer parameters and varying one parameter at time.  This analysis revealed a statistically insignificant effect of \gls{BN} kernel size (across a range from 1 to 10) based on McNemar statistical tests \cite{mcnemar1947note}.  Activation function comparisons favored \gls{ReLU} in a default configuration, but at high levels of quantization/compression showed no difference between identity and \gls{ReLU} activation functions.  

There was a clear performance loss when increasing \gls{BN} stride without a simultaneous decrease in encoder stride. We hypothesize that the model has already been optimally compressed in the time dimension (the original model has a time step of \SI{10}{ms} fed through two strides of two, resulting in an encoded frame every \SI{40}{ms}).  No dependence on encoder depth was noticeable.

In Figure~\ref{bn_results}, we see the results of varying the \gls{BNF} output dimension and quantization level at different rates of compression relative to the \SI{32} dimensional \SI{16}{bit} \gls{QMF}.  A quantization of \SI{4}{bits} and 8-12 output dimensions perform the best across compression levels.

The best performing models have been collected in Table~\ref{results_table}. Each of these models has a single hidden layer in the \gls{BNF} extractor with the exception of the 1/64 \gls{BW} model, and a stride of two in the bottleneck layer with the exception of the 1/10 \gls{BW} model. All of the models have an output quantization depth of \SI{4}{bits}, a kernel of 4, and output dimensionality between 8 and 16 channels.  They use single convolutional layer with a stride of 1 in the encoder (excepting the 1/16 and 1/32 constant time compression models, which have a stride of 2).  

Our optimized \SI{4.8}{kbps} model with a single \gls{BNF} layer actually outperforms the standard \gls{QMF} model (running at \SI{51.2}{ kbps}). Compared with the original unoptimized model, this is a \SI{160}{x} reduction in feature bandwidth for a \SI{0.6}{\%} increase in \gls{WER}.  We are able to continue to compress our \glspl{BNF} more and more heavily for slight increases in \gls{WER}. Our presented system is able to reduce the footprint of standard fixed point \gls{DSP} spectral features by a factor of 64 for a \SI{5.8}{\%} relative increase in \gls{WER}; compared with the original floating point model, this represents a \SI{960}{x} feature compression for a \SI{6.6}{\%} increase in \gls{WER}. The best performing models at \char`\~1/84 (\SI{0.6}{kbps}) and 1/128 (\SI{0.4}{kbps}) converge to \gls{WER}  values of \SI{30.36}{\%} and \SI{36.59}{\%} respectively, which represents the breakdown in performance (Figure~\ref{final_results}). 

\footnotetext[1]{These models have a reduced overall stride compared to the original model.  While the weights of the \gls{LAS} model are reduced, intermediate representations feeding the Attention model will grow \SI{2}{x} and \SI{4}{x} respectively in the time dimension. This incurs a nontrivial computational cost for the main processor, and lengthens training time.}

\section{Conclusion}

Our analysis revealed that time compression was initially the limiting factor in our model, and a \SI{40}{ms} compressed step size seems to be the limit for high accuracy models.  We found that kernel dimensionality and activation function had little effect on our results, and \SI{4}{bits} quantization with 8-12 dimensional \glspl{BNF} per timestep performed optimally.

Given these findings, we were able to design several models that effectively compress audio features on the \gls{DSP} and allow them to be cached in severely reduced memory footprints.  We designed a model that successfully compresses the original \gls{DSP} \gls{QMF} to 1/10 the size without any loss in accuracy.  As we compress the features further, we find an inflection point in \gls{WER} around \SI{1}{kbps}.    

While the models we have designed can increase the interval between main processor wake-ups by \SI{10}{x}-\SI{64}{x}, empirical data is necessary to understand the full effect on battery consumption. Some of our models require slightly more computation in the attention/decoder (because of decreased time compression), which alone may have an adverse effect on battery life. Further tuning should be done once these are tested in-situ.

These \glspl{BNF} may be useful for other compressed speech models, and the end-to-end training paradigm, while time-consuming, provides an optimal means for on-\gls{DSP} compression. We hope this architecture is adopted in portable applications as a standard technique for speech compression.  

\subsection*{Acknowledgments}

The authors would like to acknowledge Ron Weiss and the Google Brain and Speech teams for their \gls{LAS} implementation, F\'elix de Chaumont Quitry and Dick Lyon for their feedback and support, and the Google AI Z\"urich team for their help throughout the project.

\printbibliography 

\end{document}